\documentclass[11pt]{article}
\usepackage{graphicx}
\usepackage{amsfonts}
\usepackage{amssymb}
\usepackage{url}

\newcommand{\hide}[1]{}

\newtheorem{theorem}{{\bf Theorem}}

\newcommand{\ABox}{
\raisebox{3pt}{\framebox[6pt]{\rule{6pt}{0pt}}}
}
\newenvironment{proof}{{\bf Proof:}}{\hfill\ABox}

\def\S{{\mathcal S}}

{\makeatletter
 \gdef\xxxmark{%
   \expandafter\ifx\csname @mpargs\endcsname\relax % in minipage?
     \expandafter\ifx\csname @captype\endcsname\relax % in figure/caption?
       \marginpar{xxx}% not in a caption or minipage, can use marginpar
     \else
       xxx % notice trailing space
     \fi
   \else
     xxx % notice trailing space
   \fi}
 \gdef\xxx{\@ifnextchar[\xxx@lab\xxx@nolab}
 \long\gdef\xxx@lab[#1]#2{{\bf [\xxxmark #2 ---{\sc #1}]}}
 \long\gdef\xxx@nolab#1{{\bf [\xxxmark #1]}}
 \gdef\turnoffxxx{\long\gdef\xxx@lab[##1]##2{}\long\gdef\xxx@nolab##1{}}%
}

\begin{document}

\title{A New Lower Bound on Guard Placement for Wireless Localization}
\author{Mirela Damian%
   \thanks{Dept. of Computer Science, Villanova Univ., Villanova,
    PA 19085, USA.
   \protect\url{mirela.damian@villanova.edu}.}
\and
Robin Flatland%
   \thanks{Dept. of Computer Science, Siena College, Loudonville, NY 12211, USA.
    \protect\url{flatland@siena.edu}.}
\and
Joseph O'Rourke%
    \thanks{Dept. of Computer Science, Smith College, Northampton, MA
      01063, USA.
      \protect\url{orourke@cs.smith.edu}.
       %Supported by NSF Distinguished Teaching Scholars
       %award
       %DUE-0123154.
       %No longer :-(
       }
\and
Suneeta Ramaswami%
    \thanks{Dept. of Computer Science, Rutgers University,
       Camden, NJ 08102, USA.
   \protect\url{rsuneeta@camden.rutgers.edu}.}
}
\date{}
\maketitle              % typeset the title of the contribution

\begin{abstract}
The problem of \emph{wireless localization} asks to place and orient
stations in the plane, each of which broadcasts a unique key within
a fixed angular range, so that each point in the plane can determine
whether it is inside or outside a given polygonal region. The
primary goal is to minimize the number of stations. In this paper we
establish a lower bound of $\lfloor2n/3\rfloor - 1$ stations for
polygons in general position, for the case in which the placement of
stations is restricted to polygon vertices, improving upon the
existing $\lceil n/2 \rceil$ lower bound.
\end{abstract}

The problem of \emph{wireless localization} introduced
in~\cite{EGS06} asks to place a set of fixed localizers
(\emph{guards}) in the plane so as to enable mobile communication
devices to prove that they are inside or outside a secure region,
defined by the interior of a polygon $P$. The guards are equipped
with directional transmitters that can broadcast a key within a
fixed angular range. The polygon $P$ is \emph{virtual} in the sense
that it does not block broadcasts. A mobile device (henceforth, a
point in the plane) determines whether it is inside or outside $P$
from a monotone Boolean formula composed from the broadcasts using
AND($\cdot$) and OR($+$) operations only. The primary goal is to
minimize the number of guards. Solutions for convex and orthogonal
polygons were established~\cite{EGS06}, but for general polygons, a
considerable gap between a lower bound of $\lceil n/2 \rceil$ and an
upper bound of $n-2$ guards remains to be closed.
See also~\cite{o-cgc48-06}.

In this paper we establish a lower bound of $\lfloor2n/3\rfloor - 1$
guards for polygons in general position, for the case in which the
placement of guards is restricted to polygon vertices (\emph{vertex}
guards). In \cite{EGS06}, the authors use vertex guards only, and
leave open the question of whether general guards (i.e, guards
placed at arbitrary points) are more efficient. In this paper we
answer their question positively by establishing a solution with
$n/2$ general guards for a polygon that requires no fewer than
$2n/3-1$ vertex guards for localization.

A vertex guard that broadcasts over the full internal or external
angle at that vertex is called \emph{natural}. Natural guards alone
do not suffice to localize a region \cite{EGS06}, so non-natural
guards must be employed as well.

\begin{theorem}
There exist $n$-vertex simple polygons that require at least
$\lfloor 2n/3 \rfloor - 1$ guards placed at polygon vertices for
localization. \label{thm:lb}
\end{theorem}
\begin{proof}
The proof is by construction. Let $n = 3m$. Let $P$ be a polygon
consisting of $m$ narrow spikes, as illustrated in
Figure~\ref{fig:construction}. $P$ is parameterized in terms of $w$,
$h$, and $\delta$, where $\delta < h < w$. The first $m-1$ spikes
each consists of three vertices $l_i$, $t_i$, and $r_i$, for $1 \leq
i < m$. Edge $t_ir_i$ is vertical and of height $h/2$; edge
$r_il_{i+1}$ is horizontal. The vertical distance separating $l_i$
and $r_i$ is $h$; the horizontal distance between $l_i$ and $r_i$ is
$\delta$. The horizontal distance between $r_i$ and $r_{i+1}$ is
$w$. To close the polygon, the $m$th spike deviates from this
pattern slightly; its vertical edge $t_mr_m$ has height $1.5h$ and
the edge $r_ml_1$ closes the polygon.\footnote{ This polygon can be
seen as a variation on the ``comb'' polygon that establishes a lower
bound on the original art gallery problem~\cite[p.~2]{o-agta-87}
\cite[p.~6]{o-cgc-98}.
}%footnote
~We now show that $P$ cannot be localized with fewer than $2n/3-1$
guards placed at vertices.

%%%%%%%%%%%%%%%%%%%%%%%%%%%%%%%%%Figure Begin
\begin{figure}[htbp]
\centering
\includegraphics[width=.7 \linewidth]{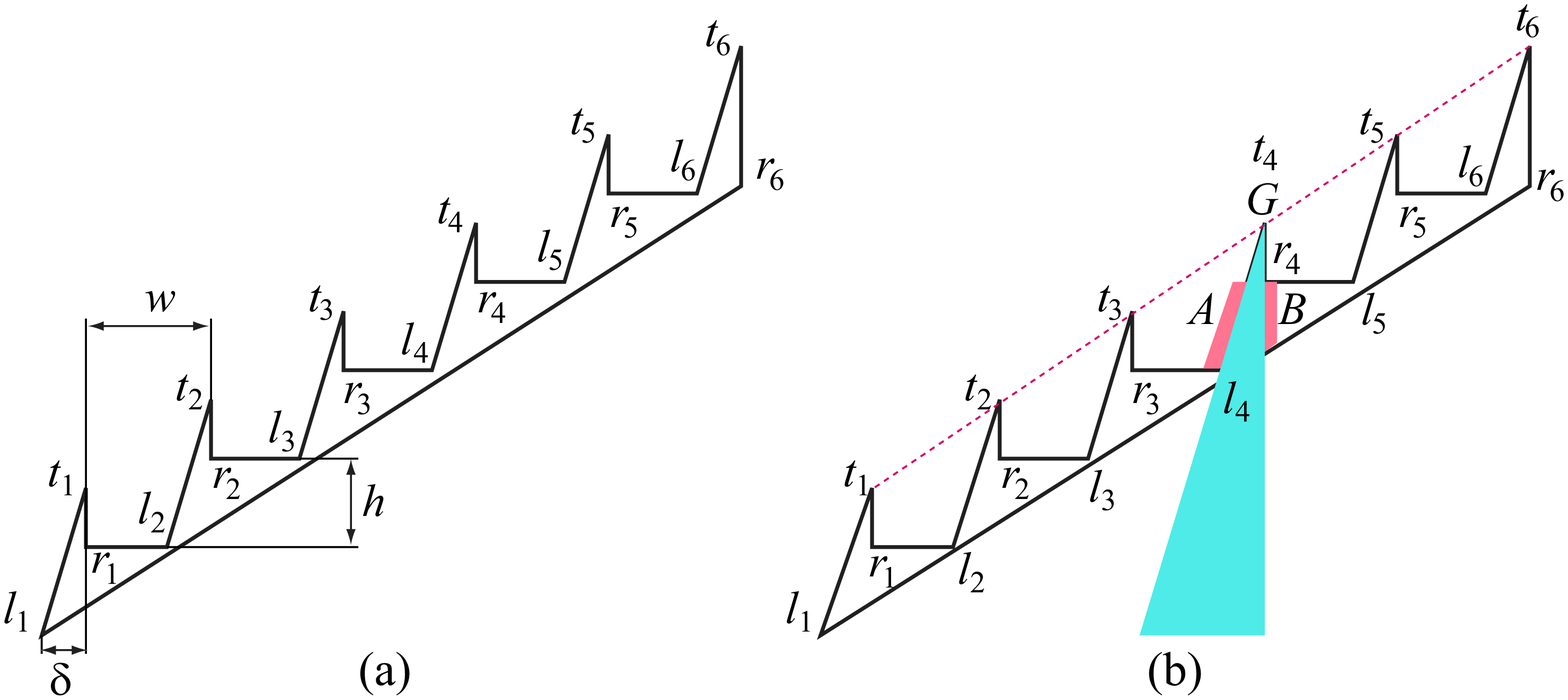}
\caption{Polygon construction.} \label{fig:construction}
\end{figure}
%%%%%%%%%%%%%%%%%%%%%%%%%%%%%%%%%Figure End

%% For each $i$, let $\ell_i$ be the left neighbor vertex of $t_i$ and
%% $r_i$ the right neighbor of $t_i$. We now list a few elementary
%% properties of $P$:
%% \begin{enumerate}
%% \item No vertex is collinear with an edge $e$ of $P$, other than the
%% endpoint vertices of $e$.
%% \item Vertices in the set $\{t_i ~|~ 1 \le i \le m\}$ are collinear,
%% and same for $\{\ell_i ~|~ 1 \le i \le m\}$ and $\{r_i ~|~ 1 \le i <
%% m\}$. The vertex $r_m$ lies slightly below the line passing though
%% $\{\ell_i ~|~ 1 \le i \le m\}$.
%% \item No point on the segment $[\ell_i,r_i]$ is visible to
%% $r_j$, for any $1 \le i, j < m$, with $j \not\in \{i-1, i\}$. Only a
%% small neighborhood of $\ell_i$ along $[\ell_i,r_i]$ is visible to
%% $r_m$, for $1 \le i < m$.
%% \item No point on the segment $(\ell_i,r_i]$ is visible to
%% $\ell_j$, for any $1 \le i, j < m$, with $j \not\in \{i, i+1\}$.
%% \end{enumerate}
%% Properties 3 and 4 above are an immediate consequence of the fact
%% that the line passing through vertices $r_i,~i = 1 \ldots m-1$, is
%% parallel to, but higher than the line passing through vertices
%% $\ell_i,~i = 1 \ldots m$ (see Fig.~\ref{fig:lattice}c).
%% Note however
%% that these properties hold even when the vertices are perturbed
%% within an $\varepsilon$-neighborhood, for small $\varepsilon > 0$,
%% so that $P$ is in general position. So from here on, we assume
%% without loss of generality that $P$ is in general position. In
%% proving the claim of the theorem, we make use of the following
%% lemma:

For any $i$, call a guard stationed at a vertex $t_i$ a \emph{tip}
guard, and a guard stationed at a vertex $\ell_i$ or $r_i$ a
\emph{base} guard. One critical observation is that each polygon
edge $e$ must align with the broadcast boundary line of a guard
$G$~\cite{EGS06}; we say that $G$ \emph{covers} $e$.
Since the only vertices in $P$ collinear with a spike edge (polygon
edges incident to $t_i$, for some $i$) are the vertices incident to
the edge,
%(cf. Property 1),
a guard covering a spike edge must be stationed at a vertex of that
edge. Counting spike edges and ignoring horizontal edges for the
moment, we get a total number of $2n/3$ spike edges that need
coverage. Next we analyze the employment of natural tip guards in an
optimal localization solution for $P$.

Let $\S$ be the set of guards in an optimal localization solution
for $P$, and let $n_0$ be the number of natural tip guards in $\S$.
The natural tip guards cover precisely $2n_0$ spike edges, leaving
$2n/3 -2n_0$ spike edges to be covered by other guards. Note however
that any other (base or non-natural tip) guard can cover at most one
spike edge (since no two spike edges are collinear).
This implies that at least $(2n/3 - 2n_0)+n_0$ guards are necessary
to cover all spike edges and therefore $|\S| \ge 2n/3 - n_0$. Thus,
if $n_0 = 0$, then $|S| \ge 2n/3$ and the proof is finished.

%%%%%%%%%%%%%%%%%%%%%%%%%%%%%%%%%Figure Begin
\begin{figure}[htbp]
\centering
\includegraphics[width=0.8\linewidth]{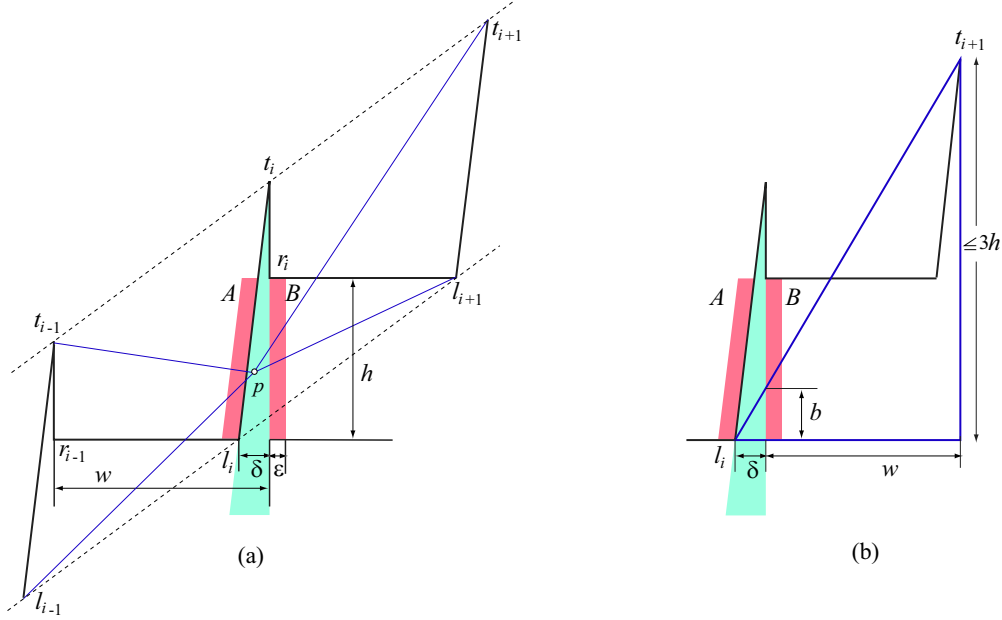}
\caption{An $A$-$B$ region requires many guards for localization.}
\label{fig:proofAB}
\end{figure}
%%%%%%%%%%%%%%%%%%%%%%%%%%%%%%%%%Figure End

Consider now the case $n_0 > 0$ and let $G$ be an arbitrary natural
tip guard at $t_i$, for $i < m$. Let $A$ and $B$ be the
$\varepsilon$-neighborhoods along the outside of $G$'s broadcast
cone between horizontal lines through $\ell_i$ and $r_i$, with $B$
restricted to the interior of $P$. See Fig.~\ref{fig:construction}b.
Observe that $G$ is only able to delineate its cone-shaped broadcast
region, leaving $A$ and $B$ with ambiguous inside/outside status.
This ambiguity can be easily resolved by a guard positioned at
$t_i$, $l_i$, or $r_i$. We show now that this is the only way to
resolve the ambiguity. Specifically, we will show that, if $A$ and
$B$ are separated by combinations of guards other than at $t_i$,
$l_i$, $r_i$, then the bound of 2n/3 is exceeded for sufficiently
small $\delta$ and sufficiently large $w$.

%Assume to the contrary that $\S$ contains no base guard adjacent to
%$G$. By construction, $|r_{i-1} r_i| = w$ and the $y$-coordinates of
%$\ell_i$ and $r_i$ differ by $h$. Let the $x$-coordinates of
%$\ell_i$ and $r_i$ differ by $\delta$ (see Fig.~\ref{fig:proofAB}a).
%One can choose $\delta$ small enough and $w$ large enough so that
%the spike becomes very thin, approaching the vertical, and the
%regions A and B become nearly $\varepsilon \times h$ rectangles.
%Moreover, since every vertex of the polygon is horizontally far away
%from $A$ and $B$, the broadcast rays crossing $A$-$B$ are nearly
%horizontal. The challenge here is to separate two regions that are
%nearly vertical, using boundary broadcast rays that are nearly
%horizontal.

First observe that any horizontal line segment $ab$, with $a \in A$
and $b \in B$, must be crossed by at least one cone edge besides
$G$'s broadcast cone edges; if this were not the case, then $a$ and
$b$ would be covered by a same set of cones and $\S$ would not
localize $P$. For any cone ray $\alpha$ not belonging to $G$, we
therefore say that its contribution to separating $A$ and $B$ is the
difference in the $y$-coordinates of the intersection points between
$\alpha$ and the cone ray boundaries for $G$. Since the height of
$A$ and of $B$ is at least $h-\delta$ (the $\delta$ term arising
because the edge $l_1r_m$ cuts off a bit of $B$), the sum of the
contributions must be at least $h-\delta$.

For any point $p$ interior to or on the boundary of the broadcast
cone for $G$, let $R_p$ denote the double-cone region bounded by the
four rays originating at $p$ and passing through $t_{i-1}$,
$\ell_{i-1}$, $t_{i+1}$ and either $\ell_{i+1}$ or $r_m$, depending
on whether $p$ lies above or below line($l_{i+1}r_m$). See
Fig.~\ref{fig:proofAB}a. Any other ray originating at $p$ and
passing through a vertex of $P$ lies inside $R_p$. Thus the
contribution of $R_p$ to separating $A$ and $B$, defined as the
maximal contribution among all such rays, is achieved by one of the
four rays bounding $R_p$. Furthermore, the contribution of $R_p$ to
separating $A$ and $B$ is maximized for $p = \ell_i$ and is achieved
by $t_{i+1}\ell_i$. This contribution value is $2.5h\delta/(\delta +
w)$, which decreases with decreasing $\delta$ and increasing $w$.

Now consider $k$ contributing rays working together to separate $A$
and $B$. It follows from the previous observations that $k > (h -
\delta)(\delta + w)/2.5h\delta$. If we choose, for instance, $\delta
= h/2$ and $w = 5nh/3$, then we get $k > 2n/3$. Thus, more than
$2n/3$ guards are required to separate $A$ and $B$ if they are
placed at vertices other than $t_i, l_i, r_i$.

So it must be that for each natural tip guard placed at a vertex
other than $t_m$, $\S$ includes an additional guard either at the
base or at the tip of the spike in order to separate regions $A$ and
$B$. Note however that such an additional guard cannot cover any
spike edges other than the ones already covered by $G$. So the total
number of guards necessary to localize $P$ is at least $n_0$ +
$(n_0-1)$ + $(2n/3 - 2n_0)$: the first term counts the natural tip
guards; the second term counts the additional guards required to
separate $A$ and $B$ for each natural tip guard (with the exception
of a natural guard placed at $t_m$); and the third term counts the
guards necessary to cover the spike edges left uncovered by the
natural tip guards. Thus at least $\lfloor 2n/3 \rfloor - 1$ guards
are necessary to localize $P$. This is also true for polygons in
general position, since the arguments here hold even when $P$'s
vertices are perturbed within an $\varepsilon$-neighborhood, for
small $\varepsilon > 0$.
\end{proof}

%%%%%%%%%%%%%%%%%%%%%%%%%%%%%%%%%Figure Begin
\begin{figure}[htbp]
\centering
\includegraphics[width=0.65\linewidth]{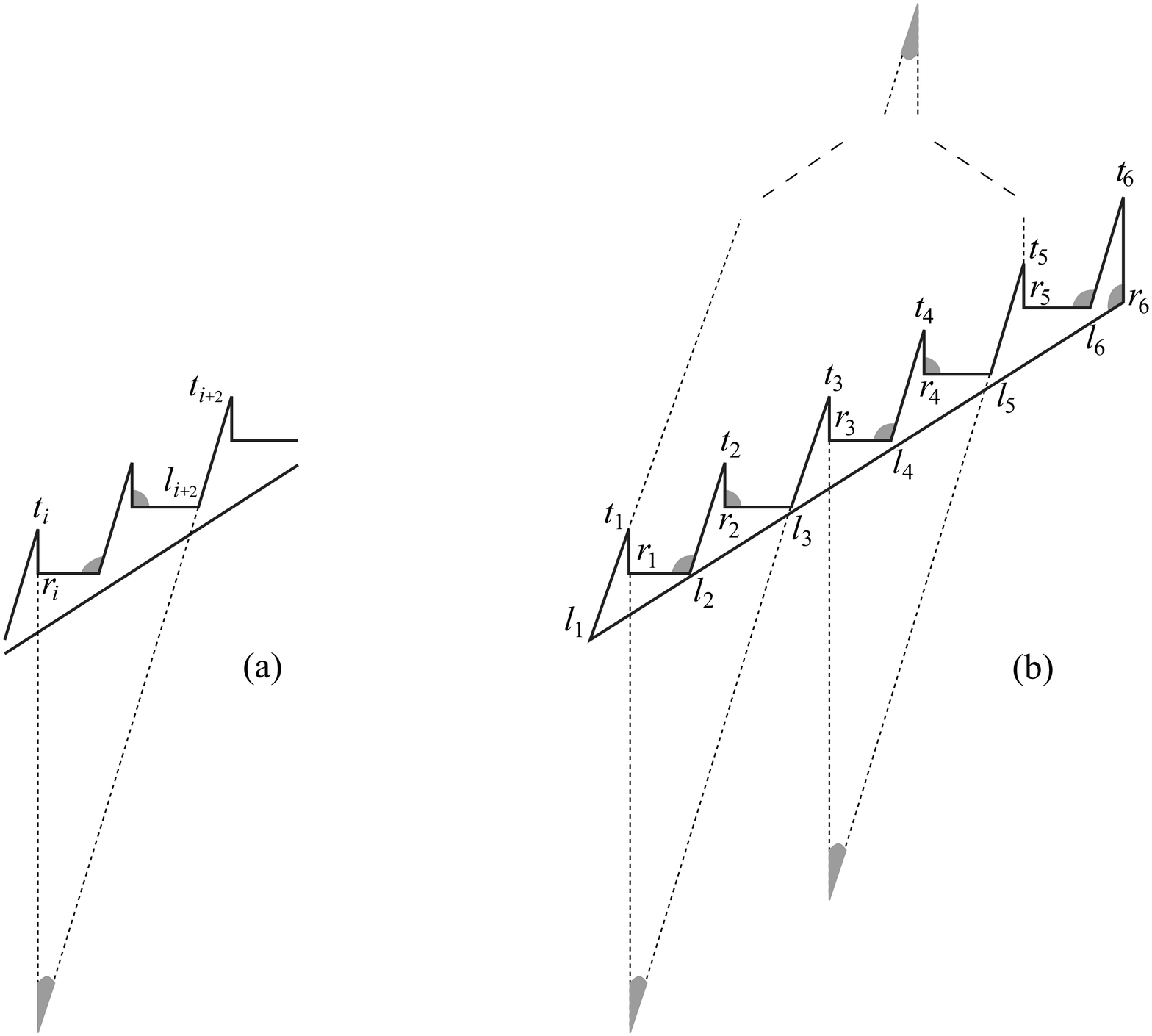}
\caption{$n/2$ general guards localize $P$.
%In the figure on the left, guards $A, B,$ and $C$ localize
%the region of the polygon to the right of $l_it_i$, above
%the horizontal line through $l_i$, and to the left of
%$l_{i+2}t_{i+2}$. This region contains two spikes.
%Polygon $P$ on the right is localized with 10 guards.
}
\label{fig:nonvertex}
\end{figure}
%%%%%%%%%%%%%%%%%%%%%%%%%%%%%%%%%Figure End
We now show that it is possible to localize the polygon $P$
constructed in Theorem~\ref{thm:lb} with $n/2$ guards, if we
eliminate the restriction that they be placed at polygon vertices,
and allow them to sit at arbitrary points. The placement of guards
is illustrated in Fig.~\ref{fig:nonvertex}. Three guards are used
for every six edges (see Fig.~\ref{fig:nonvertex}a), which implies
$n/2$ guards for $n$ edges (see Fig.~\ref{fig:nonvertex}b). In
general, if $n$ is not a multiple of $6$, then $P$ can be localized
with $\lceil n/2 \rceil+1$.

%\bibliographystyle{alpha}
%\bibliography{Guards}

\end{document}